\documentclass{pasj00}
\draft

\begin{document}
\SetRunningHead{Singh et al.}{Reliability Checks on Indo-US Library}
\Received{}
\Accepted{}

\title{Reliability checks on the Indo-US Stellar Spectral Library using Artificial
Neural Networks and Principal Component Analysis}

\author{Harinder P. \textsc{Singh}}
\affil{Department of Physics \& Astrophysics,
University of Delhi, Delhi - 110007, India}
\email{hpsingh@physics.du.ac.in}

\author{Manabu \textsc{Yuasa} and Nawo \textsc{Yamamoto}}

\affil{Research Institute for Science \& Technology,
Kinki University, Higashi-Osaka, Osaka 577-8502, Japan}
\email{yuasa@rist.kindai.ac.jp}
\and
\author{Ranjan \textsc{Gupta}}
\affil{Inter-University Center for Astronomy \& Astrophysics,
Ganeshkhind, Pune- 411007, India}\email{rag@iucaa.ernet.in}

%


\maketitle

\begin{abstract}
The Indo-US coud\'{e} feed stellar spectral library (CFLIB) made available to
the astronomical community recently by Valdes et al. (2004) contains spectra of
1273 stars in the spectral region 3460 to 9464 \AA~ at a high resolution of 1
\AA~ FWHM and a wide range of spectral types. Cross-checking the reliability
of this database is an important and desirable exercise since a number of
stars in this database have no known spectral types and a considerable
fraction of stars has not so complete coverage in the full wavelength region
of 3460-9464 \AA~ resulting in gaps ranging from a few \AA~ to several tens of
\AA~. In this paper, we use an automated classification scheme based on
Artificial Neural Networks (ANN) to classify all 1273 stars in the database. In
addition, principal component analysis (PCA) is carried out to reduce the
dimensionality of the data set before the spectra are classified by the ANN.
Most importantly, we have successfully demonstrated employment of a variation
of the PCA technique to restore the missing data in a sample of 300 stars out of the
CFLIB.

\end{abstract}

\section{Introduction}
Automated schemes of data validation and analysis have assumed added
significance recently as larger databases are increasingly
becoming available in almost all areas of observational astronomy.
With the advent of bigger CCD detectors in spectroscopy, the need for
having large libraries of stellar spectra at high spectral resolution
is also getting fulfilled. Jacoby, Hunter, \& Christian (1984, hereafter JHC)
made 158 spectra available in the range of 3510-7427\AA~ at 4.5\AA~
resolution. Prugniel \& Soubiran (2001) published a library of
708 stars using the ELODIE eschelle spectrograph at the Observatoire de
Haute-Province that covers a wavelength band of 4100-6800 \AA~ at a
resolution of R=42,000. Cenarro et al. (2001) have provided a database of 706
stellar spectra in the wavelength region 8350-9020 \AA~ at 1.5 \AA~
resolution and Le Borgne et al. (2003, hereafter STELIB) with 247 spectra in the range
of 3200-9500\AA~ at 3\AA~ resolution.
Moultaka et.al. (2004, hereafter ELODIE) compiled 1959 spectra
in the range of 4000-6800\AA~ at a resolution of 0.55\AA~.

More recently, Valdes et al. (2004) observed more than 1200 stars
with emphasis on broad wavelength coverage (3400-9500 \AA) at a resolution of
$\sim 1$ \AA~ FWHM at an original dispersion of 0.44 \AA~ per pixel. Their
coud\'{e} feed stellar spectral
library (CFLIB) provides a resolution sufficient to resolve numerous
diagnostic spectral features that can be used in the automated
parameterization of spectra.

Neural networks are a form of multiprocessor computing system, with simple
processing elements with a high degree of inter-connection, simple scalar
messaging and adaptive interaction between elements. In a supervised back
propagation algorithm, the network topology is constrained to be feed-forward,
i.e., connections are generally allowed from the input layer to
the first (and mostly only) hidden layer; from the first hidden layer
to the second,..., and from the last hidden layer to the output layer.
The hidden layer learns to recode (or to provide a representation for) the inputs.
More than one hidden layer can be used.
The architecture is more powerful than single-layer networks: it can be shown
that any mapping can be learned, given two hidden layers (of units).

Automated schemes like the Artificial Neural Networks (ANN) have been used
in Astronomy for a number of data analysis tasks like scheduling observations
(Johnson and Adorf, 1992), adaptive optics (Angel, Wizinowich and Lloyd-Hart,
1990), stellar spectral classification (Gulati et al., 1994; von Hippel et
al. 1994) and star-galaxy separation studies (Odewahn et al., 1992). In
addition Gulati, Gupta and Rao (1997) extended the ANN analysis to compare
synthetic and observed spectra of G and K dwarfs. Gulati, Gupta and Singh
(1997) estimated interstellar extinction E(B-V) using ANN from low-dispersion
ultraviolet spectra for O and B stars. Bailer-Jones, Gupta and Singh (2002)
provide a review of the ANN applications in astronomical spectroscopy.

Another powerful statistical tool for data analysis is the Principal
component analysis (PCA). It involves a mathematical procedure that
transforms a number of (possibly) correlated variables into a (smaller)
number of uncorrelated variables called principal components. The first
principal component accounts for as much of the variability in the data
as possible, and each succeeding component accounts for as much of the
remaining variability as possible. Objectives of principal component
analysis are to discover or to reduce the dimensionality of a data set and to
identify new meaningful underlying variables. The technique has been used
widely for a number of applications in Astronomy, viz., for stellar
classification by Murtagh and Heck (1987), Storrie-Lombardi et al. (1994) and
Singh et al. (1998), by Francis et al. (1992) for QSO spectra, and for
galaxy spectra by Sodr\'{e} and Cuevas (1994), Connolly et al. (1995), Lahav
et al. (1996) and Folkes, Lahav and Maddox (1996).

Another important application was developed by Unno and Yuasa (1992, 2000)
for supplementing missing observational data using a generalized PCA
technique. Subsequently, Yuasa, Unno and Magono (1999) made use of this technique
to determine distances of 183 mass-losing red giants.

The primary aim of this paper is to perform validity
checks on the CFLIB by running the ANN code on various inter library
sets like JHC, ELODIE and STELIB. In the next section we describe
the ANN analysis. In Section 3, we
demonstrate the possibility of using PCA for filling the
gaps in the spectra of CFLIB. In Section 4, we summarize important
conclusions of the study.

\begin{table}
\caption{ANN training and test cases. Two hidden layers were
used for all the cases}
\begin{center}
\begin{tabular}{cccccc}
\hline \hline
Case & Training  & Testing & $\lambda$ Region & Resolution & Class. error\\
  & (Library, & (Library, & (used) & &\\
  & No. of Spectra)& No. of Spectra) & & &\\
\hline \hline
A1& JHC, 158 & CFLIB, 1273& 4100-5500\AA~ & 4.5\AA~ & 674.3 \\
A2& JHC, 158 & ELODIE, 1959& 4100-5500\AA~ & 4.5\AA~ & 514.8 \\
A3& JHC, 158 & STELIB, 247& 3600-7400\AA~ & 4.5\AA~ & 861.2 \\
B1& ELODIE, 174 & ELODIE, 1959 & 4000-5500\AA~ & 1\AA~ & 496.4 \\
B2& ELODIE, 174 & ELODIE, 1959 & 4000-6800\AA~ & 1\AA~ & 576.0 \\
B3& ELODIE, 174 & CFLIB, 1272 & 4000-5500\AA~ & 1\AA~ & 742.7 \\
B4& ELODIE, 174 & CFLIB, 1273 & 4000-6800\AA~ & 1\AA~ & 848.5 \\
C1& STELIB, 247 & CFLIB, 1273 & 4000-6800\AA~ & 3\AA~ & 643.9\\
C2& STELIB, 247 & CFLIB, 1273 & 3500-9400\AA~ & 3\AA~  & 670.6\\
C3& STELIB, 247 & ELODIE, 1959 & 4000-6800\AA~ & 3\AA~ & 501.8\\
\hline \hline
\end{tabular}
\end{center}
\end{table}

\begin{figure}
\begin{center}
\FigureFile(120mm,120mm){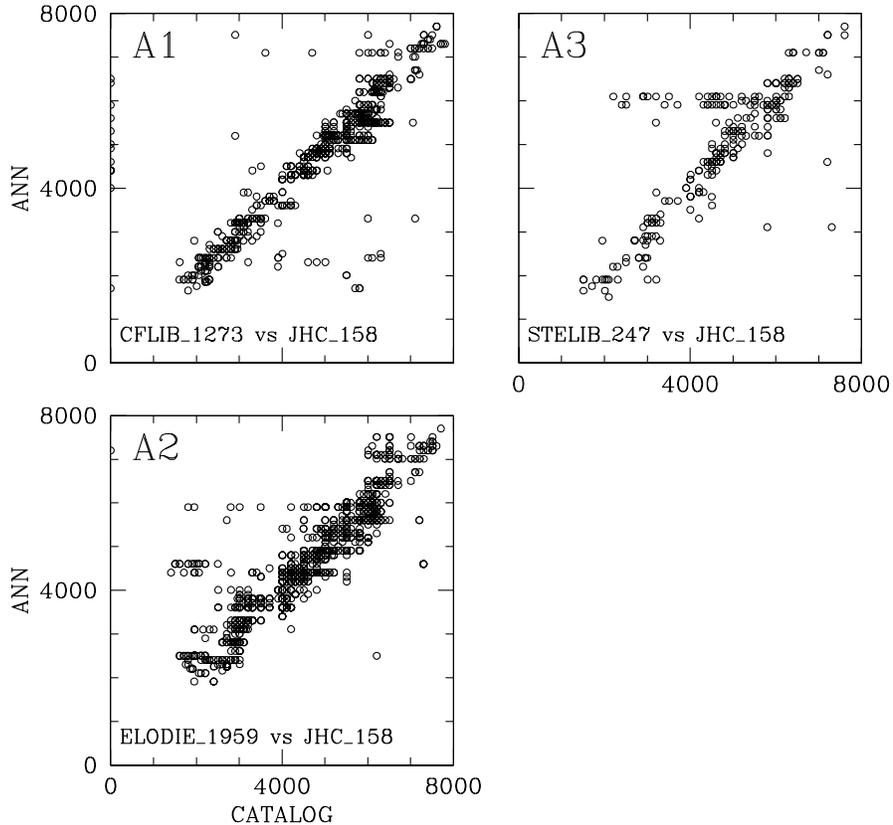}
\end{center}
\caption{Classification scatter plots for Cases A1-A3}
\end{figure}

\section{ANN Analysis}

A classification scheme using ANN involves two stages. A training stage and a
testing stage. In the training stage, the input patterns and the desired
output patterns are defined before the learning process of the ANN is carried
out. During training, the network output and the desired output are compared
and the network weights are adjusted. We employ a back propagation algorithm
(Rumelhart, Hinton and Williams, 1986) to achieve this. The learning is
stopped when the desired error threshold is reached and the network weights
are frozen for use with the test set. In the testing stage, the test patterns
are used by the network and classified in terms of the training classes.

In what follows, we have trained the ANN on three libraries with a view to
run validity checks on the CFLIB. Three libraries that we used for training
are:

1. JHC (Jacoby, Hunter and Christian, 1984) with 158 spectra available
in the range of 3510-7427\AA~ at 4.5\AA~ resolution

2. ELODIE (Moultaka et al., 2004) with 1959 spectra
in the range of 4000-6800\AA~ at a resolution of 0.55\AA~

3. STELIB (Le Borgne et al., 2003) with 247 spectra in the range
of 3200-9500\AA~ at 3\AA~ resolution

\begin{figure}
\begin{center}
\FigureFile(120mm,120mm){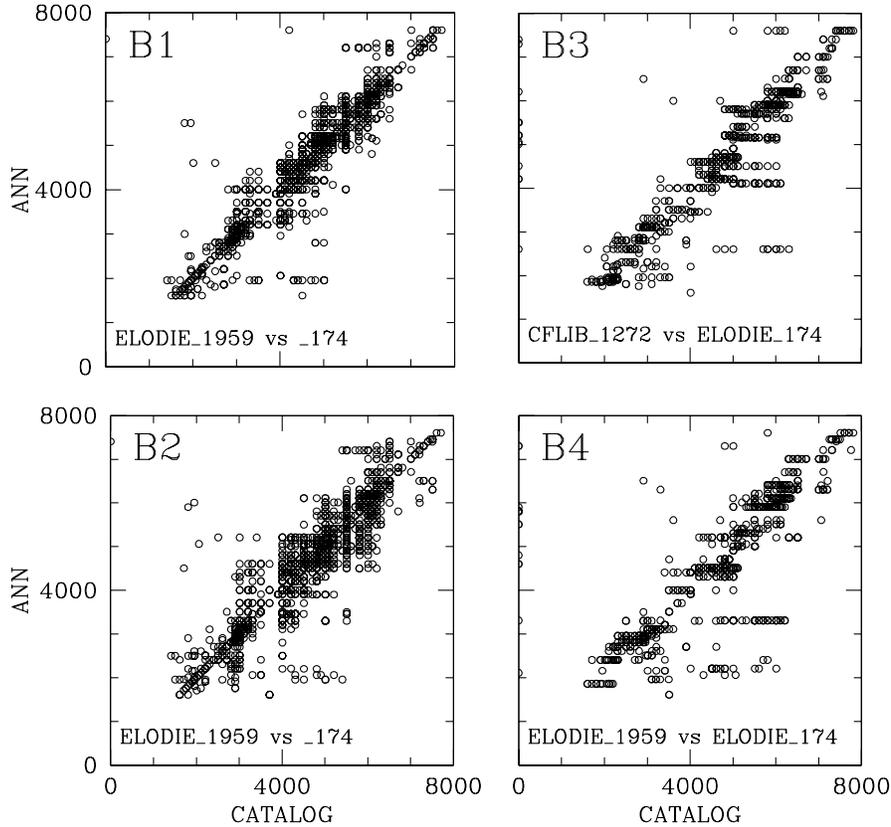}
\end{center}
\caption{Classification scatter plots for Case B1-B4}
\label{fig:sample}
\end{figure}

A total of 10 test cases were run and are summarized in Table 1.
We also give the wavelength region used in each case, resolution and the
resulting classification error. Cases A1-A3 involved training of the ANN by
JHC and testing on, respectively, the CFLIB, ELODIE, and STELIB. Cases B1-B4
involved training the ANN with 174 individual classes of ELODIE and testing
on ELODIE (4000-5500 \AA~, Case B1), ELODIE (4000-6800 \AA~, Case B2), CFLIB
(4000-5500 \AA~, Case B3), and CFLIB (4000-6800 \AA~, Case B4). The last
three cases C1-C3 involved training on STELIB and testing on CFLIB (4000-6800 \AA~, Case
C1), CFLIB (3500-9400 \AA~, Case C2), and ELODIE (4000-6800 \AA~, Case C3).

The resolution used for training and testing for cases A1-A3 is 4.5 \AA~
which is the resolution of JHC, 1 \AA~ for cases B1-B4 which is the
resolution of CFLIB, and 3 \AA~ for cases C1-C3 which is the resolution of
STELIB. The higher resolutions of some of the testing cases were degraded to match
the resolution of the training library by gaussian convolution.
The testing is done on the complete libraries, i.e., for 247 stars of
STELIB, 1959 stars of ELODIE, and 1273 stars of CFLIB.

The classification scatter plots from these analysis are shown in
 Figures 1-3. The MK spectral types have been coded numerically for
use by the ANN. The MK alphabetic class O is given a numeric code of 1000, B
is 2000, A is 3000, and so on with M being 7000. The subclasses are
multiplied by 100 and added in, thus a F5 star is coded as 4500. The
luminosity types I, II, III, IV, and V are coded as 1.5, 3.5, 5.5, 7.5, and
9.5, respectively. A F5V star is thus coded as 4509.5. In the present scheme,
the luminosity class is given a low weight and the main emphasis on classification
of the spectral types.

\begin{figure}
\begin{center}
\FigureFile(120mm,120mm){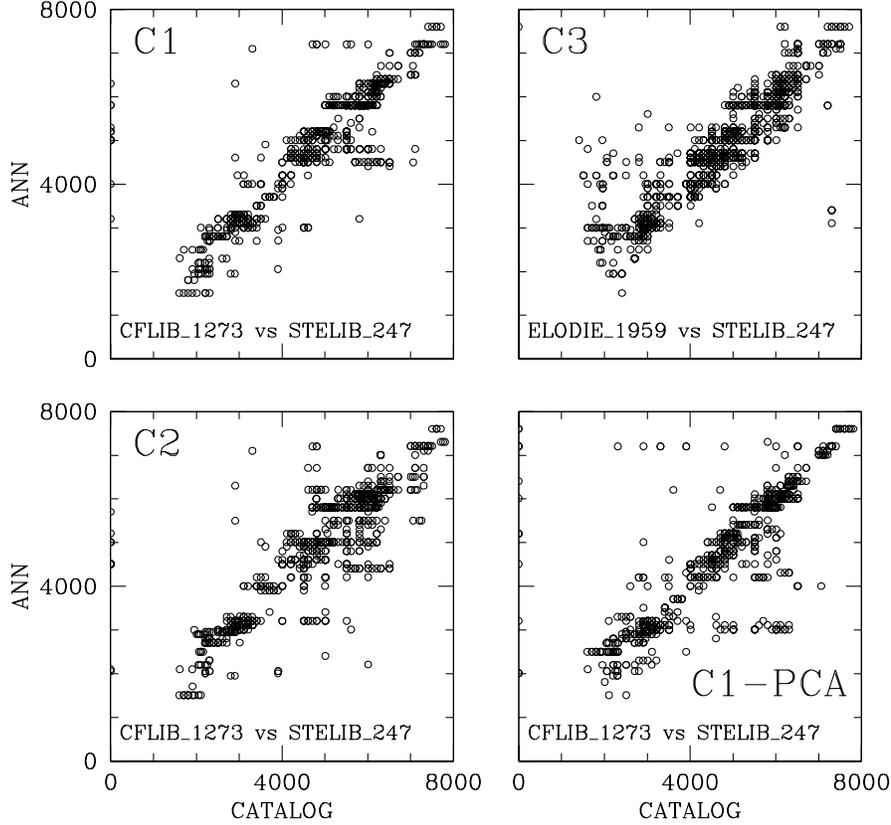}
\end{center}
\caption{Classification scatter plots for Case C1-C3. Also shown is case C1
with PCA preprocessing where first 15 principal components were used}
\label{fig:sample}
\end{figure}

\begin{figure}
\begin{center}
\FigureFile(120mm,120mm){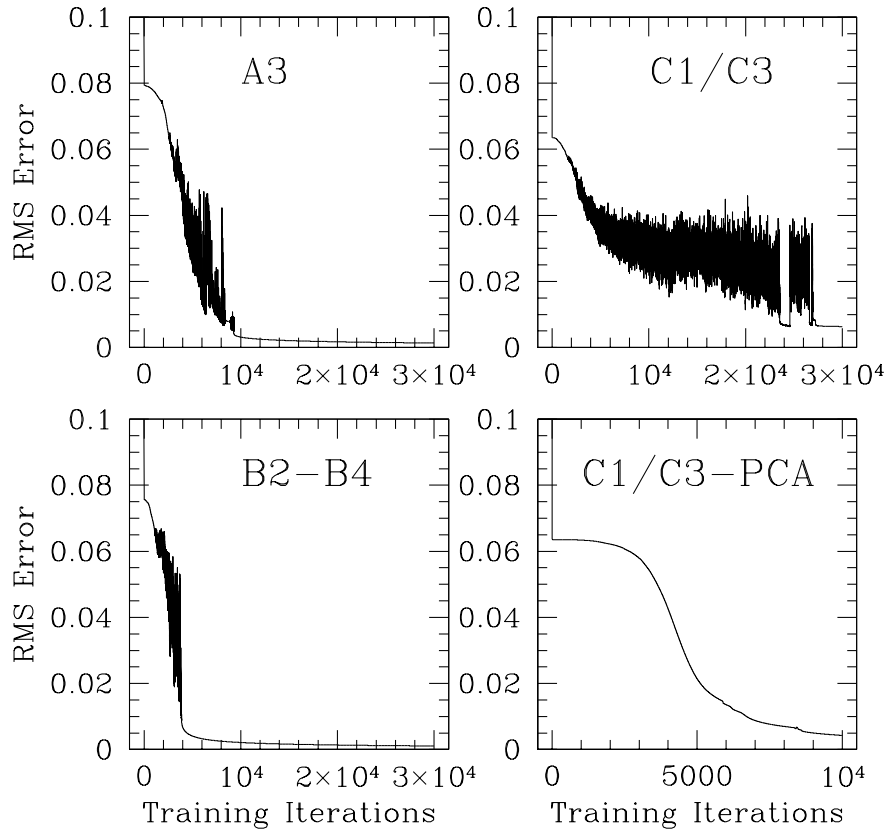}
\end{center}
\caption{ANN training session learning curves for some selected
cases}
\label{fig:sample}
\end{figure}

In the following we provide a discussion on the classification accuracy of
the three sets of cases:

\noindent {\it Cases A1-A3}

The Cases A1-A3 use JHC as the training set with A1 and A2 in the wavelength
region 4100-5500\AA~ i.e. in the blue part of the spectrum. In the case A3,
JHC was trained for the full span i.e. 3600-7400\AA~ and tested on STELIB
in the same band. The blue region classification error is lowest for
ELODIE and somewhat inferior for CFLIB. However, the full span
classification of STELIB deteriorates the error to about 861 i.e. 8.6
sub-spectral-type.

\noindent {\it Cases B1-B4}

In Case B1-B4, ELODIE was used as the training set and was tested on both
ELODIE and CFLIB in the blue and full span. The training set of 174
spectra was preselected  from amongst the  1959 ELODIE full set with one example
spectra per spectro-luminosity class. The best classification
is obtained for the case B1 with blue region of ELODIE. For the
same region of CFLIB, i.e., case B2, the classification is somewhat
inferior. The full span classification case B3 for ELODIE
and case B4 for CFLIB are consequently with higher classification errors.

\noindent {\it Cases C1-C3}

Cases C1-C3 use STELIB as the training set and the test sets are
CFLIB limited span C1; ELODIE limited span C3 and CFLIB full span
case C2. The limited span cases of C1 and C3 show errors of 6.43 and 5.01
sub-spectral types but remarkably the case C2 with full span shows
an error of only 6.7.

The most useful result as far as CFLIB is concerned, is the case C2
since the largest wavelength span is used for this case.
It may be noted that except JHC, all other spectral libraries have gaps
in them and this contributes greatly into the classification errors.
Further, we also used a PCA based preprocessor for all the cases
to reduce the dimensionality of the train-test sets as described
in Singh et al. (1998). The resulting classification errors using first 15
principal components are only marginally poorer than the ones listed in the
last column of Table 1. Last panel in Fig. 4 shows the scatter plot of case
C1 with PCA.

Figure 4 displays some example learning curves of the ANN training
sessions for Cases A3, B2-B4, C1-C3 and a PCA version of C1-C3.
The B2-B4 (and C1-C3) are same training sessions.
All the learning curves fall to a low rms learning error at the level
on number of iterations of 30,000. The PCA version however requires
much less number on iterations ($\sim$ 10000) to bring down these
errors.

\begin{figure}
\begin{center}
\FigureFile(80mm,80mm){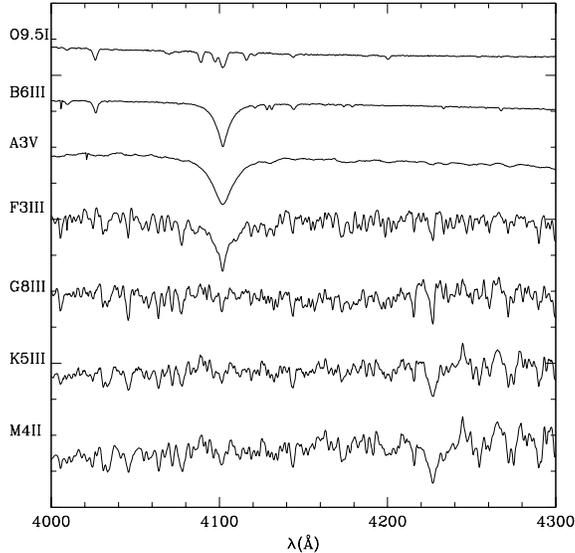}
\end{center}
\caption{Representative spectra of seven stars out of the total
of 300 stars. The spectral types are listed on the vertical axis}
\label{fig:sample}
\end{figure}

\section{Restoration of Missing Data using PCA}

To fill the gaps in the spectra one can use the data
from a similar type star which doesn't
have a gap at the same wavelength. However,
we present here a preliminary study of an automated method for
restoration of missing
data for a set of spectra of 300 stars in the wavelength region
4000-4300 \AA~ selected from the CFLIB.
The list of stars with their spectral types
is given in Table 3. The method of restoration
is adopted from Unno and Yuasa (1992) and is briefly described here
for the present data set.

To begin with, we have 301 flux values at 1 \AA~ interval in the
range for 4000-4300 \AA~ for 300 stars. A sample of seven spectra
are shown in Figure 5. For the i-th star, let
$F_j^i$ and $w_j^i$ be the j-th observed flux value and its weight
respectively, where j = 1,....., n (n = 301) and i = 1,....., N (N = 300).
If a particular flux value $F_j^i$ for a particular star is missing, its
weight is equal to zero.

For applying the PCA, the normalized data $f_j^i$ is defined as
\begin{equation}
f_j^i = \frac{[F_j^i\,-\,<F_j>]}{\sigma_j},
\end{equation}

\noindent where $<F_j>$ and $\sigma_j$ are the mean and the standard deviation
of $F_j$ respectively and are given by

\begin{equation}
<F_j> = \frac{\sum_{i=1}^{N} w_j^i F_j^i}{\sum_{i=1}^{N} w_j^i},
\end{equation}

\noindent and
\begin{equation}
\sigma_j^2 = \frac{\sum_{i=1}^{N} w_j^i [F_j^i - <F_j>]^2}{\sum_{i=1}^{N}
w_j^i}.
\end{equation}

Following Unno and Yuasa (1992), we define virtual
data $x_j^i$ and their corresponding weight $v_j^i$ for each observed flux value
$f_j^i$ for each star as
\begin{eqnarray}
v_j^i & = & 1 - w_j^i,\\
\sum_{i=1}^{N} v_j^i x_j^i & = & 0,\,\,\,\,
\sum_{i=1}^{N} v_j^i (x_j^i)^2 = \sum_{i=1}^{N}v_j^i.
\end{eqnarray}

Equation (5) represents the statistical constraint that the mean value of
the virtual data is zero and the standard deviation is unity. For such a
case, the correlation coefficient between the j-th quantity and the k-th
quantity is defined by
\begin{equation}
C_{jk} = \frac{1}{N} \sum_{i=1}^{N}(w_j^i f_j^i + v_j^i x_j^i)(w_k^i f_k^i +
v_k^i x_k^i).
\end{equation}
The most probable value of $x_j^i$ are thus given by the following set
of n simultaneous linear algebraic equations:
\begin{equation}
\sum_{l=1}^{n}\frac{1}{\lambda_l}\left[\mu_{lj}^2 x_j^i + \sum_{k \neq j}\mu_{lj}\mu_{lk}
(w_k^i f_k^i + v_k^i x_k^i)\right] = 0,\,\,\,\,\, (j = 1,......, n),
\end{equation}

\noindent where $\lambda_l$ is the $l$-th eigenvalue and $\mu_l^j$ represent
the $j$-th component of the $l$-th eigenvector in the PCA.
The final adjusted value for the normalized $j$-th flux value for the $i$-th star
is given by
\begin{equation}
w_j^i f_j^i + v_j^i x_j^i.
\end{equation}

To check the veracity of this procedure let us  assume that from the
data set only one flux value $F_1^s$ is missing. This means that $w_1^s = 0$
and all the other weights $w_j^i$ except $w_1^s$ are equal to unity. In this
simplified situation, equations (7) reduce to
\begin{equation}
\left(\sum_{l=1}^n \frac{\mu_{l1}^2}{\lambda_l}\right) x_1^s +
\sum_{l=1}^n \frac{\mu_{l1}}{\lambda_l} \left(\sum_{k=2}^n \mu_{lk} f_k^s\right)
= 0,\,\,\,\,\,(s=1,.....,N)
\end{equation}

Equation (9) can be easily solved to get $x_1^s$ for the missing flux
$f_1^s$. By exchanging the columns, one can compute $x_2^s$, $x_3^s$, and
so on for the missing flux for any wavelength and for any star. Six separate
cases were studied and are listed in Table 2. The results
of our preliminary analysis are shown in Figures 6 and 7.

\begin{table}
\caption{Cases studied for flux restoration analysis using PCA}
\begin{center}
\begin{tabular}{cccc}
\hline \hline
Case & Flux  & No. of Principal & $\lambda$\\
     & Reconst. at & Components & Region\\
\hline \hline
(a) & 4000 \AA~ & 10 & 4000-4009 \AA~\\
(b) & 4077 \AA~ & 10 & 4077-4086 \AA~\\
(c) & 4291 \AA~ & 10 & 4291-4300 \AA~\\
(d) & 4000 \AA~ & 20 & 4000-4019 \AA~\\
(e) & 4077 \AA~ & 20 & 4077-4096 \AA~\\
(f) & 4281 \AA~ & 20 & 4281-4300 \AA~\\
\hline \hline
\end{tabular}
\end{center}
\end{table}

\begin{figure}
\begin{center}
\FigureFile(120mm,120mm){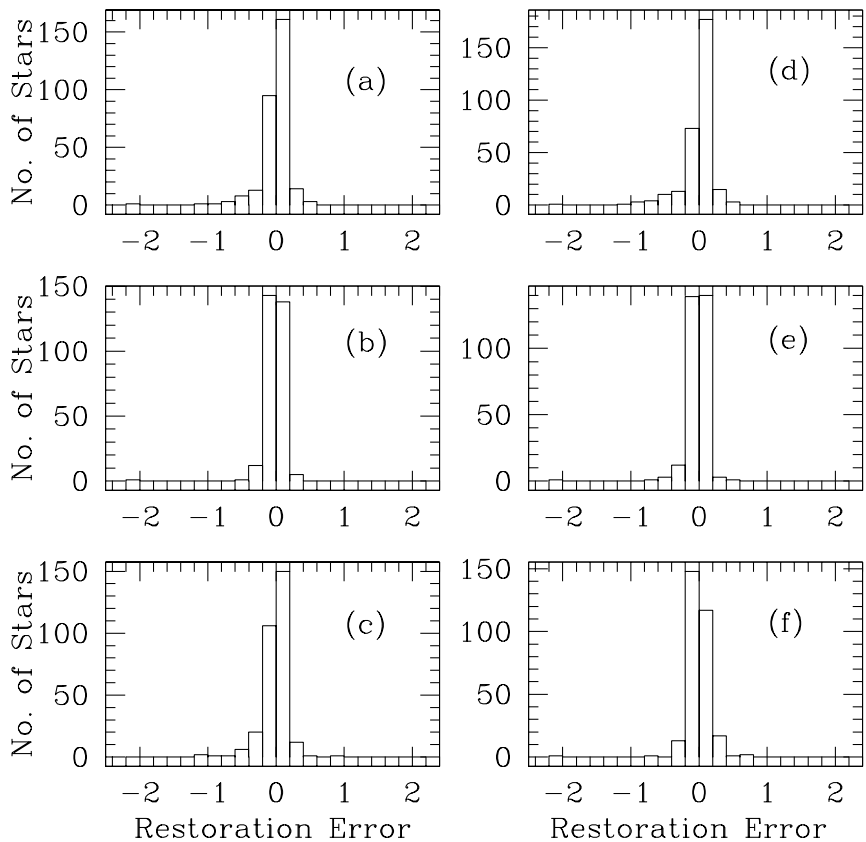}
\end{center}
\caption{Histogram of restoration error and the number of stars for the
six cases. All stars with restoration errors greater than $\pm 0.2$ are
binned together}
\end{figure}

Case (a)uses a flux region of 10 \AA~ starting from 4000 \AA~ and thus 10
principal components to reconstruct the fluxes at 4000 \AA~ for all the 300
stars. From Fig. 6(a) we see that 256 stars have restoration error ($f_1^i -
x_1^i$) within $\pm 0.1$. Fig. 7(a) shows the restored error vs. the original
value of the normalized flux $f_1^i$ (with mean zero). it is clear that for
most of the stars the reconstruction is good. Case (d) uses 20 principal
components and the reconstruction of flux at 4000 \AA~ is comparable, maybe
slightly worse as is clear from Fig. 6(d) and Fig. 7(d).

\begin{figure}
\begin{center}
\FigureFile(120mm,120mm){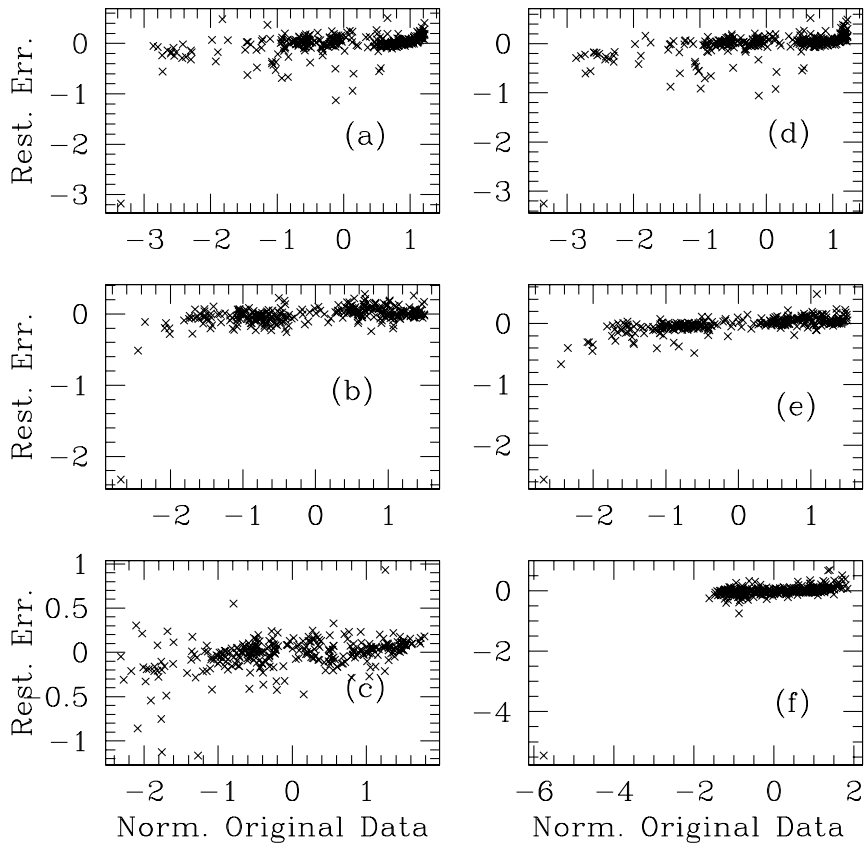}
\end{center}
\caption{Restoration error plotted against the normalized original data
for all the six cases. The outlier at the bottom left of most of the plots
is HD 31996}
\end{figure}

We also tested the validity of this flux reconstruction method by attempting
to reproduce a strong absorption feature, namely the 4077 \AA~ SrII feature,
visible in the representative spectra in Fig. 5. Case (b) attempts to achieve
this using 10 principal components rather successfully as is clear from
Fig. 6(b) and Fig. 7(b) respectively. The flux at 4077 \AA~ was
reconstructed to within $\pm 0.1$ for a total of 281 stars out
of 300. The case (e) with 20 principal components
also reproduces the flux at 4077 \AA~ well although no better than case (b)
(279 stars within $\pm 0.1$)
suggesting that the 10 PC's are enough for the data reconstruction in this
data set. Lastly, cases (c) and (f) show similar behaviour in reconstructing
towards the end of the wavelength interval and the results are
plotted in Fig. 6(c and f) and Fig. 7 (c and f).

Another interesting offshoot of this analysis was in picking outliers,
stars which have either no known MK spectral types or have noisy spectra.
Fig. 7(a,b,d,e,f) clearly show one outlier for which our scheme is
unable to restore or reconstruct the fluxes. The star is HD 31996 and it
indeed has no MK spectral class assigned to it (Table 3) as was verified from
the CFLIB and the SIMBAD database. Another star, HD 46687, also has
no known spectral type but its spectrum resembles a M type star.
For this star, our analysis was able to reconstruct the fluxes.
The spectra of these two stars obtained from the CFLIB are plotted
in Fig. 8.

\begin{figure}
\begin{center}
\FigureFile(80mm,80mm){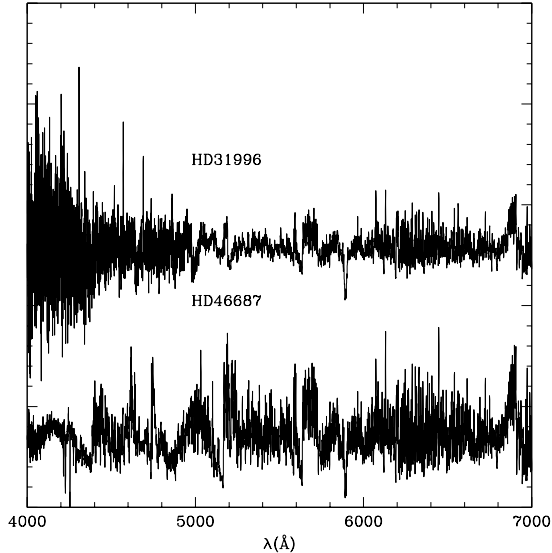}
\end{center}
\caption{Spectrum of the two stars HD 31996 and HD 46687 which have no known
spectral types. Star HD 31996 is the outlier in the flux restoration analysis}
\end{figure}

\section{Conclusions}

We have performed an extensive analysis based on artificial neural networks
to classify stars in an automated manner in the Indo-US CFLIB using three
databases viz., JHC, STELIB and ELODIE. The main aim of this exercise was
two-fold. One was to perform the reliability checks on CFLIB to see how the
gaps in the library affect the classification accuracy. We find that the
despite the presence of gaps, we have achieved classification accuracy of
less than one main class. The second aim was to evolve and test automated
procedures of classifying stellar spectra. This was achieved by trying our
ANN scheme on ten different cases of training and testing on different
pairs of libraries. The schemes are numerically intensive, with the ANN
training stage requiring several hours of CPU time on the fastest of
workstations. A PCA analysis was employed successfully to reduce the
dimensionality of the data set and hence faster training without appreciable
loss in classification accuracy.

Both STELIB and ELODIE libraries have variable spans of wavelength gaps where the
fluxes are filled with zeros (similar to CFLIB). Such
gaps lead to classification errors in the PCA and ANN schemes. However, we
have carried out some preliminary analysis with the basic $\chi^2$ minimization
scheme on these libraries wherein the gap portions were omitted in both
train and test sets. This lead to a remarkable improvements in the
classification accuracy.

We have also employed a generalized principal component analysis to first
create and then fill the gaps in a sample of 300 stars out of the CFLIB
in the blue region. At present, we have used a simplified system to
reconstruct flux values at one wavelength bin at a time for these 300 stars.
We hope to exploit the full potential of the scheme and attempt to fill
larger gaps in stellar spectra in a subsequent study.

MY and HPS are grateful to JSPS (Japan Society for Promotion of Science)
and DST (Department of Science \& Technology, India) for financial support
for exchange visits which made this work possible.
MY would like to thank Emeritus Professor W. Unno of the
University of Tokyo for helpful discussions. The research has made use
of the SIMBAD database, operated by CDS, Strasbourg, France and the INDO-US CFLIB
managed by NOAO, Tucson, AZ, USA.


\appendix
\section{List of stars for data restoration analysis}
The list of 300 stars with their HD numbers and the spectral types is given
in Table 3. Two stars HD 31996 and HD 46687 have no known spectral types.

\begin{longtable}{llllll}
  \caption{Sample of 300 stars used for data restoration analysis}
  \hline\hline
   HD No. & Sp. Type & HD No. & Sp. Type & HD No. & Sp. Type  \\
\endfirsthead
  \hline\hline
   HD No. & Sp. Type & HD No. & Sp. Type & HD No. & Sp. Type  \\
  \hline \hline
\endhead
  \hline
\endfoot
  \hline
\endlastfoot
100889 & B9.5V  & 141680  & G8III  &        191615  & G8IV \\
102212 & M1III  & 141714  & G3.5III  &      195324  & A1I \\
102224 & K0.5III  &       142091  & K1IV  & 196867  & B9IV \\
102328 & K3III  & 142198  & K0III  &        198001  & A1V \\
102870 & F9V  &   142373  & F8V  &   25329  & K1V \\
103047 & K0  &    143107  & K2III  &         28978  & A2V \\
103287 & A0V  &   143666  & G8III  &         29613  & K0III \\
104985 & G9III  & 143761  & G0V  &   29645  & G0V \\
105043 & K2III  & 145328  & K1III  &         30562  & F8V \\
105262 & B9  &    146791  & G9.5III  &       30614  & O9.5I \\
106365 & K2III  & 147677  & K0III  &         30652  & F6V \\
106714 & G8III  & 148387  & G8III  &         30739  & A1V \\
107113 & F4V  &   148783  & M6III  &         30743  & F5V \\
107213 & F8V &   148786  & K0III  &         30812  & K1III \\
107383  & G8III  &        149161  & K4III  &         31295  & A0V \\
107418  & K0III  &        149630  & B9V  &   31421  & K2III \\
107950  & G6III  &        149661  & K2V  &   31996 &   0.0\\
108225  & G9III  &        149757  & O9V  &   32147  & K3V \\
108954  & F9V &  150012  & F5IV  &  33111  & A3III \\
109317  & K0III  &        150100  & B9.5V  &         33256  & F2V \\
109345  & K0III  &        150117  & B9V  &   35468  & B2III \\
109358  & G0V  &  150449  & K1III  &         35497  & B7III \\
110281  & K5  &   150453  & F3V  &   36673  & F0I \\
110897  & G0V  &  150680  & G0IV &  36861 &  O8III \\
111335  & K5III  &        150997  & G7.5III &       37043  & O9III \\
111591  & K0III  &        151431  & A3V  &   37160  & K0III \\
111765  & K4III  &        151613  & F2V  &   37984  & K1III \\
111812  & G0III  &        151769  & F7IV &  38656  & G8III \\
112300  & M3III  &        151862  & A1V  &   38899  & B9IV \\
113226  & G8III  &        152569  & F0V  &   39003  & G9.5III \\
113436  & A3V  &  152614  & B8V &   39283  & A2V \\
113848  & F4V  &  152815  & G8III  &         39587  & G0V \\
113996  & K5III  &         15318  & B9III  &         39801 &  M1 \\
114038  & K1III &        153597  & F6V  &   39853 &  K5III \\
114092  & K4III  &        153653  & A7V &   39866  & A2II \\
114330  & A1IV  & 153808  & A0V  &   40035  & K0III \\
114357  & K3III  &        154278  & K1III  &         40111  & B0.5II \\
114642  & F6V  &  154431  & A5V &   40136  & F1V \\
114710  & F9.5V  &        154445  & B1V  &   40183 &  A2IV \\
115004  & K0III  &        154660  & A9V  &   40239  & M3II \\
115136  & K2III  &        155763  & B6III  &         40536  & A6 \\
115202  & K1III  &        157741  & B9V  &   40801  & K0II \\
115383  & G0V  &  157910  & G5III  &         41117  & B2I \\
115539  & G8III  &        158716  & A1V  &   41330  & G0V \\
115604  & F3III &        158899  & K4III  &         41597  & G8III \\
115617  & G5V  &  159332  & F6V  &   41636  & G9III \\
116292  & K0III  &        160765  & A1V  &   41692 &  B5IV \\
116656  & A2V  &  161056  & B1.5V  &         42475  & M1I \\
117176  & G5V  &  161868  & A0V  &   42543  & M1I \\
117243  & G5III  &        163588  & K2III  &         43039  & G8.5III \\
117818  & K0III  &        163917  & G9III  &         43042  & F6V \\
117876  & G8III  &        163989  & F6IV  &  43232  & K1.5III \\
118055  & K0  &   164259  & F2IV  &  43247  & B9II \\
118266  & K1III  &        164284  & B2V  &   43318  & F6V \\
120136  & F6IV & 164353  & B5I  &   43380  & K2III \\
120164  & K0III &        165029  & A0V  &   43827  & K1III \\
120348  & K1III  &        165358  & A2V &   43947  & F8V \\
120452  & K0III  &        165401  & G0V  &   44007  & G5IV \\
121146  & K2IV & 165645  & F0V  &   44033  & K3I \\
121370  & G0IV  & 165687  & K0III  &         44478  & M3III \\
122563  & F8IV  & 165760  & G8III  &         44537  & M0I \\
123657  & M4.5III  &      165908  & F7V  &   44769  & A5IV \\
123977  & K0III  &        166014  & B9.5V &         44951  & K3III \\
124570  & F6IV & 166046  & A3V  &   45282  & G0 \\
124850  & F7IV  & 166207  & K0III  &         45410  & K0III \\
124897  & K1.5III  &      167042  & K1III  &         45412  & F8I \\
125451  & F5IV  & 168151  & F5V  &   46184  & K1III \\
125454  & G8III &        168656  & G8III  &         46687  & 0.0\\
126141  & F5V  &  168723  & K0III  &         47105  & A0IV \\
126271  & K4III  &        169191  & K3III  &         47205  & K1III \\
126868  & G2IV  & 169414  & K2III  &         47731  & G5I \\
127334  & G5V  &  170693  & K1.5III  &       47839  & O7V \\
128000  & K5III  &        171301  & B8IV  &  48329  & G8I \\
128750  & K2III &        171391  & G8III  &         48432  & K0III \\
129312  & G7III  &        172569  & F0V  &   48433  & K1III \\
129336  & G8III  &        173087  & B5V  &   48737  & F5IV \\
129956  & B9.5V  &        173399  & G5IV  &  48781  & K1III \\
129972  & G8.5III  &      175317  & F6IV  &  50420  & A9III \\
129978  & K2III  &        175535  & G7III  &         51309  & B3I \\
130948  & G1V  &  175545  & K2III  &         54662  & O7III \\
131111  & K0III  &        175588  & M4II  &  54719  & K2III \\
131156  & G8V  &  175640  & B9III  &         55280  & K2III \\
132132  & K1III  &        175743  & K1III &         55575  & G0V \\
132345  & K3III  &        175751  & K2III &         57264  & G8III \\
133165  & K0.5III  &      176301  & B7III &         57669  & K0III\\
133208  & G8III  &        176318  & B7IV &  57727  & G8III\\
134083  & F5V  &  176582 &  B5IV  &  58207  & G9III \\
134190  & G7.5III  &      176819 &  B2IV  &  58343  & B2V \\
135742  & B8V &  177724 &  A0V  &   58551  & F6V \\
136064  & F9IV  & 177817 &  B7V &   59881  & F0III \\
136202  & F8III  &        178125 &  B8III  &         60179  & A1V \\
136512  & K0III  &        178329  & B3V  &   61064  & F6III \\
136726  & K4III  &        180006  & G8III  &         61295  & F6II \\
137052  & F5IV  & 180711  & G9III  &         62509  & K0III \\
138716  & K1IV  & 182293  & K3IV  &  63302  & K3I \\
139195  & K0III  &        182568 &  B3IV &  65714  & G8III \\
139446  & G8III  &        183144 &  B4III  &         65900  & A1V \\
139641  & G7.5III  &      184915 &  B0.5III  &       67228  & G1IV \\
140027  & G8III  &        188350  & A0III  &         69897  & F6V \\
141004  & G0V  &  191243  & B5I  &   70110  & F9V \\
\end{longtable}


\end{document}